\documentclass[12pt]{article} 
\usepackage[top=1.25in, bottom=1.25in, left=1in, right=1in]{geometry}
\usepackage[sectionbib]{natbib}
\usepackage{array,epsfig,fancyheadings,rotating}
\usepackage{fancyhdr}
\usepackage{titling}
\usepackage[]{hyperref}  
\usepackage{sectsty, secdot}
\sectionfont{\fontsize{12}{14pt plus.8pt minus .6pt}\selectfont}
\renewcommand{\theequation}{\thesection\arabic{equation}}
\subsectionfont{\fontsize{12}{14pt plus.8pt minus .6pt}\selectfont}

\usepackage{amsmath}
\usepackage{amssymb}
\usepackage{amsfonts}
\usepackage{multirow}
\usepackage{amsthm}
\usepackage{algorithm}

\theoremstyle{definition}

\usepackage{color}

\author{Ma T. and Ren Z.}

\begin{document}


\markright{ \hbox{\footnotesize\rm Statistica Sinica
}\hfill\\[-13pt]
\hbox{\footnotesize\rm
}\hfill }

\markboth{\hfill{\footnotesize\rm FIRSTNAME1 LASTNAME1 AND FIRSTNAME2 LASTNAME2} \hfill}
{\hfill {\footnotesize\rm Tianzhou Ma and Zhao Ren} \hfill}

\renewcommand{\thefootnote}{}
$\ $\par


\fontsize{12}{14pt plus.8pt minus .6pt}\selectfont \vspace{0.8pc}
\centerline{\large\bf Robust distance correlation for variable screening}
\vspace{.4cm} \centerline{Tianzhou Ma$^1$, Hongjie Ke$^1$ and Zhao Ren$^2$} \vspace{.4cm} \centerline{\it
$^1$ University of Maryland  \\ $^2$ University of Pittsburgh} \vspace{.55cm} \fontsize{9}{11.5pt plus.8pt minus
.6pt}\selectfont


\begin{quotation}
\noindent {\it Abstract:}

High-dimensional data are commonly seen in modern statistical applications, variable selection methods play indispensable roles in identifying the critical features for scientific discoveries. Traditional best subset selection methods are computationally intractable with a large number of features, while regularization methods such as Lasso, SCAD and their variants perform poorly in ultrahigh-dimensional data due to low computational efficiency and unstable algorithm. Sure screening methods have become popular alternatives by first rapidly reducing the dimension using simple measures such as marginal correlation then applying any regularization methods. A number of screening methods for different models or problems have been developed since the seminal work by Fan \& Lv (2008), however,   none of the methods have targeted at data with heavy-tailedness, which is another important characteristics of modern big data. In this paper, we propose a robust distance correlation (``RDC'') based sure screening method to perform screening in ultrahigh-dimensional regression with heavy-tailed data. The proposed method shares the same good properties as the original model-free distance correlation based screening while has additional merit of robustly estimating the distance correlation when data is heavy-tailed and improves the model selection performance in screening. We conducted extensive simulations under different scenarios of heavy tailedness to demonstrate the advantage of our proposed procedure as compared to other existing model-based or model-free screening procedures with improved feature selection and prediction performance. We also applied the method to high-dimensional heavy-tailed RNA sequencing (RNA-seq) data of The Cancer Genome Atlas (TCGA) pancreatic cancer cohort and RDC was shown to outperform the other methods in prioritizing the most essential and biologically meaningful genes.

\vspace{9pt}
\noindent {\it Key words and phrases:}
Distance correlation, Robustness, Sure screening property, Variable selection, Truncation
\par
\end{quotation}\par

\def\thefigure{\arabic{figure}}
\def\thetable{\arabic{table}}

\renewcommand{\theequation}{\thesection.\arabic{equation}}

\fontsize{12}{14pt plus.8pt minus .6pt}\selectfont

\setcounter{section}{1} 
\setcounter{equation}{0} 
\noindent {\bf 1. Introduction}
\label{sec:intro}

Data with much larger number of features than sample size is frequently seen in modern statistical applications, ranging from genomic research, biomedical imaging to signal processing problems. High dimensional variable selection has become fundamentally important to many scientific discoveries nowadays. For example, oncologists apply variable selection methods to identify genes most predictive of cancer prognosis from tens of thousands of candidate genes. Classical variable selection methods such as the best subset selection using AIC or BIC criterion suffers from an NP-hard combinatorial problem and the computation becomes unaffordable with a large number of features. Since late 1990s, various regularization methods have been developed and extensively applied for feature selection in high-dimensional regression and classification models. These methods include but are not limited to, Lasso \citep{tibshirani1996regression}; SCAD \citep{fan2001variable,kim2008smoothly,zou2008one}; the least angle regression (LARS) algorithm \citep{efron2004least}; elastic net \citep{zou2005regularization,zou2009adaptive}; the adaptive Lasso \citep{zou2006adaptive}; the Dantzig selector \citep{candes2007dantzig} and MCP \citep{zhang2010nearly}. One can refer to \cite{fan2010selective} for a detailed overview of variable selection methods in high-dimensional models. As the number of features $p$ grows exponentially with the sample size $n$ in an ultrahigh-dimensional problem, however, many regularization methods perform poorly owing to the low computational efficiency and unstable algorithm. In a seminal work, \cite{fan2008sure} proposed a sure independence screening (SIS) method by first ranking and selecting features based on their marginal correlations with the response in a linear model before applying any regularization methods. SIS is attractive in practice by rapidly reducing the dimension in the original model while keeping the most important features in the reduced set with large probability (a property known as ``sure screening property''). From then on, variable screening has drawn considerable attention and a wide variety of screening methods have been developed for different models or problems in the past decade. \cite{fan2009ultrahigh} and \cite{fan2010sure} developed more general version of independent screening by ranking the maximum marginal likelihood estimators for generalized linear model. Other papers have proposed sure screening methods for nonparametric or semi-parametric regression \citep{fan2011nonparametric,liu2014feature,fan2014nonparametric,chang2016local}, for quantile regression  \citep{he2013quantile,wu2015conditional,ma2017variable}, for censoring data \citep{fan2010high,zhao2012principled,gorst2013independent,song2014censored}, for Gaussian graphical model \citep{luo2014sure,liang2015equivalent}, for classification problems \citep{mai2012kolmogorov,cui2015model,pan2016ultrahigh,xie2019category} and among others. In addition to those methods targeting at a class of specific models, \cite{zhu2011model}, \cite{li2012feature} and \cite{mai2015fused} proposed model-free screening procedures for a more general usage when the model structure cannot be clearly specified.

A majority of regularization methods and marginal correlation based screening methods assumed light-tailed distribution for response and covariates \citep{fan2008sure,fan2010selective}. The violation of such an assumption is common in practice. For instance, non-Gaussian and heavy-tailed distribution is a stylized feature of high-dimensional data such as microarray and next-generation sequencing (NGS) data in genomics and functional magnetic resonance imaging (fMRI) data in neuroimaging. For example, the gene expression measured by either bulk RNA-seq or single-cell RNA-seq techniques display both zero inflation and heavy tail behavior that approximates a power law \citep{townes2020quantile}. To tackle such a problem, numerous robust regularization methods have been developed in the literature based on quantile regression or least absolute regression \citep{li20081,wu2009variable,wang2013l1,fan2014adaptive}. However, these methods are essentially targeting at median estimation instead of mean estimation. In a deviation analysis, \cite{catoni2012challenging} showed that the empirical mean is way deviated in the presence of heavy-tailed errors. By revisiting the classical Huber loss as a way of robustification \citep{huber1964robust}, \cite{fan2017estimation}, \cite{zhou2018new} and \cite{sun2019adaptive} proposed Huber-type estimators of regression coefficients and derived their non-asymptotic concentration results in a linear model under mild moment conditions on the error. In these methods, a tuning parameter (called robustification parameter) needs to be specified beforehand to balance between robustness and bias of estimation. Similar Huber estimator has been proposed for estimating high-dimensional covariance matrix with the robustification parameter determined via empirical cross-validation \citep{avella2018robust} or in a data-driven fashion \citep{ke2018user}. However, no methods have been developed for the robust estimation of measures (e.g. Pearson correlation or distance correlation) directly used in variable screening when heavy-tailed distributions are observed. 

First introduced by \cite{szekely2007measuring} and then used as a utility for variable screening in \cite{li2012feature}, distance correlation (DC) is a new measure of dependence and has impressive properties that makes it useful in feature screening and stand out among all screening procedures. Specifically, distance correlation is robust to model misspecification and equivalent to Pearson correlation in screening when normality assumption holds. In addition, it measures the dependence between two random vectors so it can be readily used for multivariate responses or group wise predictors often seen in real data. Later work on DC considers bias correction in DC estimation \citep{szekely2013distance} and extension from marginal to partial DC \citep{szekely2014partial}. Despite its robustness against model specification, the sample DC statistics used in \cite{li2012feature} lacks tail robustness. In this paper, we propose a DC-based measure via a truncation-type robust estimation (called ``Robust DC'') for sure screening in ultrahigh-dimensional regression with heavy-tailed data. The proposed Robust DC shares the same good properties as the original distance correlation based screening when data is subexponential. In addition, it has the merit of robustly estimating the distance correlation when data is heavy-tailed and improves the model selection performance in screening. The proposed measure enjoys the sure screening property with weaker assumptions on the distribution than the original distance correlation screening, as justified by our theoretical analysis. Our theorem also reveals a new phenomenon on the optimal choice of optimal robustification parameter. To the best of our knowledge, the proposed method is the only screening method that considers both robustness to model specification and tail robustness. We conducted extensive simulations under different scenarios of heavy tailedness to demonstrate the advantage of our proposed procedure as compared to other existing model-based or model-free screening procedures. The proposed screening procedure was also applied to the high-dimensional heavy-tailed RNA sequencing (RNA-seq) data of The Cancer Genome Atlas (TCGA) pancreatic cancer cohort. Robust DC was shown to outperform the other methods in prioritizing the most essential and biologically meaningful genes and improving both the feature selection and prediction performance.

The rest of the paper is organized as follows. In Section \ref{sec:method}, we will first introduce the distance correlation measure and the adaptive Huber type method, and then propose our robust distance correlation measure for feature screening in ultra-high dimensional regression. Section \ref{sec:theory} provides the non-asymptotic theoretical results of the proposed measure and shows the sure screening property. We conduct extensive simulations in Section \ref{sec:sim} to show the strength of our method. In Section \ref{sec:real}, we apply our method to an example of RNA-seq study in pancreatic cancer. We conclude and discuss the potential extension of the method in Section \ref{sec:discuss}.

\setcounter{equation}{0} 
\section{Methods}
\label{sec:method}

\subsection{Independence screening using distance correlation}
\label{subsec:pre}

\cite{szekely2007measuring} first introduced the DC measure to generalize the idea of correlation to characterize the joint dependence of any two random vectors. Denote by $\phi_{\mathbf{x}}(t)$ and $\phi_{\mathbf{y}}(s)$ the respective characteristic functions of two random vectors $\mathbf{x}$ and $\mathbf{y}$, and $\phi_{\mathbf{x},\mathbf{y}}(t,s)$ be the joint characteristic function. The nonnegative distance covariance between $\mathbf{x}$ and $\mathbf{y}$ with finite first moments is defined as: 
$$ dcov^2(\mathbf{x},\mathbf{y})= \int_{R^{d_x+d_y}}\frac{|| \phi_{\mathbf{x},\mathbf{y}}(t,s) -  \phi_{\mathbf{x}}(t) \phi_{\mathbf{y}}(s) ||^2} {c_{d_x} c_{d_y} ||t||^{1+d_x}_{d_x}  ||s||^{1+d_y}_{d_y} }  dtds , $$  
where $d_x$ and $d_y$ are the dimensions of $\mathbf{x}$ and $\mathbf{y}$, $c_d=\pi^{(1+d)/2}/\Gamma((1+d)/2)$, and $||\mathbf{a}||_d$ indicates the Euclidean norm of $\mathbf{a} \in R^d$.
 Like the Pearson correlation, the nonnegative DC between $\mathbf{x}$ and $\mathbf{y}$ with finite first moments is defined:
$$ dc(\mathbf{x},\mathbf{y}) = \frac{ dcov(\mathbf{x},\mathbf{y})}{\sqrt{ dcov(\mathbf{x},\mathbf{x}) dcov(\mathbf{y},\mathbf{y}) } }. $$

Further, \cite{szekely2007measuring} showed that:  
$$dcov^2(\mathbf{x},\mathbf{y}) = S_1 + S_2 - 2S_3, $$
where $S_1$, $S_2$ and $S_ 3$ are defined as: $S_1 = E\{|| \mathbf{x} - \mathbf{x'} ||_{d_x}  || \mathbf{y} - \mathbf{y'} ||_{d_y}  \}, $ 
$ S_2 = E\{|| \mathbf{x} - \mathbf{x'} ||_{d_x}\} E\{ || \mathbf{y} - \mathbf{y'} ||_{d_y}  \} \text{ and} $
$ S_3 = E\{E(|| \mathbf{x} - \mathbf{x'} ||_{d_x} |\mathbf{x} ) E( || \mathbf{y} - \mathbf{y'} ||_{d_y}  |\mathbf{y})\}, $
$(\mathbf{x'}, \mathbf{y'})$ is an independent copy of $(\mathbf{x}, \mathbf{y})$. 

This has motivated the estimation of sample distance covariance via estimating $S_1$, $S_2$ and $S_3$ using the usual moment estimation:
$$ \widehat{dcov}^2(\mathbf{x},\mathbf{y}) = \hat{S}_1 + \hat{S}_2 - 2\hat{S}_3, $$ 
where $\hat{S}_1 = \frac{1}{n^2}\sum\limits_{k=1}^n\sum\limits_{l=1}^n ||\mathbf{x}_k - \mathbf{x}_l ||_{d_x} ||\mathbf{y}_k - \mathbf{y}_l ||_{d_y}  $, $\hat{S}_2 = \frac{1}{n^2} \sum\limits_{k=1}^n\sum\limits_{l=1}^n  ||\mathbf{x}_k - \mathbf{x}_l ||_{d_x} \frac{1}{n^2} \sum\limits_{k=1}^n\sum\limits_{l=1}^n ||\mathbf{y}_k - \mathbf{y}_l ||_{d_y}  $ and $\hat{S}_3 =\frac{1}{n^3}\sum\limits_{k=1}^n \sum\limits_{l=1}^n \sum\limits_{m=1}^n ||\mathbf{x}_k - \mathbf{x}_l ||_{d_x} ||\mathbf{y}_m - \mathbf{y}_l ||_{d_y} $. When $d_x=d_y=1$, they are simplified to $\hat{S}_1 = \frac{1}{n^2}\sum\limits_{k=1}^n\sum\limits_{l=1}^n |X_k - X_l ||Y_k - Y_l |$, $\hat{S}_2 = \frac{1}{n^2} \sum\limits_{k=1}^n\sum\limits_{l=1}^n  |X_k - X_l | \frac{1}{n^2} \sum\limits_{k=1}^n\sum\limits_{l=1}^n |Y_k - Y_l | $ and $\hat{S}_3 =\frac{1}{n^3}\sum\limits_{k=1}^n \sum\limits_{l=1}^n \sum\limits_{m=1}^n |X_k - X_l | |Y_m - Y_l |$. Accordingly, the sample DC can be estimated as 
\begin{equation} \label{eqn:sdc}
 \widehat{dc}(\mathbf{x},\mathbf{y})  = \frac{\widehat{dcov}(\mathbf{x},\mathbf{y}) }{ \sqrt{ \widehat{dcov}(\mathbf{x},\mathbf{x})  \widehat{dcov}(\mathbf{y},\mathbf{y})  } }.
 \end{equation}

DC has three important properties that make it a preferred measure for variable screening: (1) DC is monotonically increasing in Pearson correlation under normality assumption, thus using DC for screening is equivalent to using Pearson correlation for linear regression; (2) DC is equal to zero if and only if $\mathbf{x}$ and $\mathbf{y}$ are independent, thus DC is more effective than Pearson correlation when there exist non-linear relationship between two random variables; (3) DC measures the dependency between two random vectors so it allows for both multivariate responses ($d_y>1$) and groupwise predictors ($d_x>1$), either continuous or categorical.  

Without loss of generality, consider the case with univariate response, denote by $Y$ the univariate response and $\mathbf{x}=(X_1,\ldots, X_p)^T$ the predictor vector in a regression, where $p$ is the number of features. For ultrahigh-dimensional regression, the dimensionality $p$ grows at an exponential rate of sample size $n$. \cite{li2012feature} proposed to use $\hat{\omega}_j = \widehat{dc}^2(X_j,Y)$ as a measure in independence screening of ultrahigh-dimensional regression. They selected important features with large $\hat{\omega}_j$ as follows: $\{j: \hat{\omega}_j \ge cn^{-\kappa}, 1\le j\le p \}$, where $c$ and $\kappa$ are pre-specified threshold values theoretically defined. They also showed the sure screening property of their procedure by assuming both $x$ and $y$ have the subexponential tails. In practice, they followed from \cite{fan2008sure} to set the empirical threshold and recommended the first $d=[n/\log n]$, $2[n/\log n]$ or $3[n/\log n]$ features with the largest $\hat{\omega}_j$ to be retained after screening. 

\subsection{Huber loss and robust distance correlation for independence screening}
\label{subsec:Rdc}

Heavy-tailedness is one prominent feature of high-dimensional data that has brought new challenges to conventional statistical methods. In the presence of heavy-tailed distribution, there is a non-negligible probability that observations are sampled far away from the population mean causing the empirical mean to seriously deviate from the true mean \citep{catoni2012challenging}. To tackle heavy-tailed errors, most methods in the literature conducted least absolute deviation (LAD) regression, but they essentially alter the problem by estimating the median or quantiles instead of the mean. In most cases, however, the mean estimate but not the median is the quantity of interest, e.g. in the estimation of DC. 

Alternatively, one can consider the Huber loss as another way of robustification to adapt for different magnitudes of errors. The Huber loss $ l_\tau(.)$ \citep{huber1964robust} is defined as:
\[  l_\tau(x) = \left\{
\begin{array}{ll}
        x^2/2, & \qquad \text{if } |x| \le \tau, \\
       \tau |x| - \tau^2/2, & \qquad \text{if } |x| > \tau, \\
\end{array} 
\right. \]
where $\tau>0$ is called the robustification parameter. Two extremes of the Huber loss are the least squares and the LAD as $\tau=\infty$ and $\tau=0$ respectively. When a majority of $|x|$'s are small, it simplifies to the common squared error loss. In the presence of heavy-tailed errors, the Huber loss down-weights large $|x|$ (i.e. potential outliers) to introduce robustness. In general, the minimizer becomes more robust but also more biased from the model parameters as $\tau$ decreases, so the robustification parameter $\tau$ controls the tradeoff between bias and robustness. Choosing $\tau$ to achieve an optimal tradeoff between two has become the major task in the development of Huber-type robust methods.

\cite{fan2017estimation} considered high-dimensional regression with coefficients in an $\ell_q$ ball ($0<q\leq 1$) and proposed a modified Huber's procedure to obtain a robust estimator of coefficients with the robustification parameter adapted to the sample size and dimension. \cite{sun2019adaptive} further proposed the adaptive Huber regression for robust estimation and inference, allowing $q=0$ and more general moment condition of the error in both low and high dimensions.  In a similar vein, \cite{ke2018user} directly considered a simple truncated estimator $\psi_\tau(x)=(|x|\wedge \tau)sign(x)$ for covariance estimation and also proposed simple automatic data-driven tuning scheme (see also \cite{wang2018new}) to find the optimal robustification parameter for the computation of the robust estimators.  

In spite of the nice properties and being model-free, the sample DC estimate in equation (\ref{eqn:sdc}) used in \cite{szekely2007measuring} and \cite{li2012feature} is not robust against heavy-tailedness of the data. When the underlying distribution is heavy-tailed, the observations might be sampled far away from the population mean. As a result, the individual sample DC estimate will not be concentrated around the truth which directly impacts its screening performance, given that there are many DC estimates to be considered. To guarantee the tail-robustness in the estimation of DC, we propose a robust estimates of DC (called ``Robust DC'' or RDC in short) via obtaining the truncated estimators $\hat{S}^{\tau_1,\tau_2}_1$, $\hat{S}^{\tau_1,\tau_2}_2$ and $\hat{S}^{\tau_1,\tau_2}_3$. Suppose we have $n$ i.i.d. samples of two random variables of interest $Y$ and $X$, the robust covariance and correlation estimates are defined as: 
\begin{equation}
\label{eq:1}
 \begin{split}
\widehat{rdcov}^2(Y,X) = \hat{S}^{\tau_1,\tau_2}_1 + \hat{S}^{\tau_1,\tau_2}_2 - 2\hat{S}^{\tau_1,\tau_2}_3;  \\ 
 \widehat{rdc}(Y,X)  = \frac{\widehat{rdcov}(Y,X) }{ \sqrt{ \widehat{rdcov}(Y,Y)  \widehat{rdcov}(X,X)  } }, \\
 \end{split}
\end{equation}  
where $\hat{S}^{\tau_1,\tau_2}_1 = \frac{1}{n^2}\sum\limits_{k=1}^n\sum\limits_{l=1}^n |\psi_{\tau_1}(Y_k - Y_l) | |\psi_{\tau_2}(X_k - X_l) |  $, $\hat{S}^{\tau_1,\tau_2}_2 = \frac{1}{n^2} \sum\limits_{k=1}^n\sum\limits_{l=1}^n  |\psi_{\tau_2}(X_k - X_l) | \frac{1}{n^2} \sum\limits_{k=1}^n\sum\limits_{l=1}^n | \psi_{\tau_1}(Y_k - Y_l) | $ and $\hat{S}^{\tau_1,\tau_2}_3 =\frac{1}{n^3}\sum\limits_{k=1}^n \sum\limits_{l=1}^n \sum\limits_{m=1}^n | \psi_{\tau_1}(Y_k - Y_l) | | \psi_{\tau_2}(X_m - X_l) |$. The truncation operator $\psi_\tau(u)= (|u|\wedge \tau)sign(u)$, and $\tau$ is the robustification parameter. Interestingly, the optimal choice of $\tau$ for DC estimate is revealed by a careful theoretical investigation: it should be adapted to the sample size $n$ at the rate of $n^{1/4}$ rather than the typical rate $n^{1/2}$ shown in existing works. In other words, we should truncate the data more aggressively in calculating the robust estimates of distance correlation. Details on the theoretical properties of optimal choice of $\tau$ and how to determine it in a data-driven fashion will be discussed in Section \ref{subsec:tune} and xx.

With the predictor vector $\mathbf{x}=(X_1,\ldots, X_p)^T$ and the response variable $Y$, we will use $\hat{\omega}^{\tau}_j = \widehat{rdc}^2(X_j, Y)$ as a marginal utility to rank the relative importance of $X_j$ and perform independence screening in ultrahigh-dimensional regression. The features with the largest $\hat{\omega}^{\tau}_j$ are kept after screening, i.e. the set $\{j: \hat{\omega}^{\tau}_j \ge c'n^{-\kappa'}, 1\le j\le p \}$, where $c'$ and $\kappa'$ are theoretically justified threshold values. In practice, as for all screening methods, the top number of features to be left $d$ is a key tuning parameter that empirically balances the tradeoff between false negative (when $d$ is too small) and false positive (when $d$ is too large) errors. In this paper, we followed from the literature and chose $d=\lfloor n/\log n \rfloor$. But we also suggest users to try out the alternative approaches by linking $d$ to the significance level of a hypothesis test \citep{buhlmann2010variable} or the desired false positive rate \citep{zhao2012principled} for the best performance. 

\noindent \textbf{Remarks:} Multivariate response or grouped predictors are common to see in the modern statistical applications. For example, many complex diseases consist of a large number of highly correlated clinical phenotypes that are associated with pleiotropic genes or genetic variants. \cite{li2012feature} showed in simulations that the DC can be successfully used to screen for multivariate response or grouped predictors. Though we introduced the method using univariate response and predictor notation, our method is built on the basis of the original DC thus is readily extended to perform robust screening for multivariate response (when $d_y>1$) or grouped predictors (when $d_x>1$). The tuning of robustification parameter will remain the same in the multivariate case. 

\subsection{Tuning of the robustification parameter}
\label{subsec:tune}
A data-driven procedure is proposed to automatically tune the robustification parameter $\tau$. Define the pairwise difference by a generic notation $\{Z_1,\ldots, Z_N \} = \{|Y_1 - Y_2|, \ldots, |Y_{n-1} - Y_n| \}$ for the response $Y$ or $\{|X_{1j} - X_{2j}|, \ldots, |X_{(n-1)j} - X_{nj} | \}$ for the $j$th predictor of $\mathbf{x}$, where $N={n \choose 2}$. By theorem xx, it can be shown that an ``ideal'' choice of $\tau$ can be solved from the following equation: 
\begin{equation}
\label{eq:2}
 \frac{1}{N}\sum\limits_{i=1}^N \frac{(Z_i^4 \wedge \tau^4 ) }{\tau^4} = \frac{t}{n},
 \end{equation}  
where $t=C \log p$ as guided by the theory. The choice of $C$ is the same for all $X_j$'s ($1\le j\le p$) and $Y$ and can be determined empirically depending on the extent of tail heaviness of the data. From the above, we can get an optimal estimate of $\tau$ for the response and each of the predictor variable, respectively. We then plug $\tau$ estimates back into equation \ref{eq:1} to compute each of $\hat{S}^{\tau_1,\tau_2}_1$, $\hat{S}^{\tau_1,\tau_2}_2$ and $\hat{S}^{\tau_1,\tau_2}_3$ and finally solve for $\widehat{rdcov}(X_j,Y)$ and $\widehat{rdc}(X_j,Y)$.



\subsection{Algorithm for RDC based sure independence screening (RDC-SIS)}
\label{subsec:algo}

Below is an algorithm to estimate RDC and perform sure screening based on its ranking: 
 \begin{algorithm}
   \caption{RDC-SIS}
  \begin{itemize}
    \item[] \textbf{Input}: Predictor vectors $\mathbf{x}_i = (X_{i1},\ldots,X_{ip})^T \in R^p $ and the outcome variable $Y_i$ for $i=1,\ldots,n$; pre-specified $C$ and $d$. 
   \item[]\textbf{Estimation}: Data-adaptive estimation of RDC
         \begin{itemize}  
      \item[]1: Calculate the pairwise difference of each predictor variable $\{Z_{j1},\ldots, Z_{jN}\}$, $1\leq j\leq p$ and the outcome variable $\{Z_{y1},\ldots, Z_{yN}\}$.
      \item[]2: Given $t=C \log p$, solve $\tau_j$ and $\tau_y$ for each predictor variable and the outcome variable respectively from equation (\ref{eq:2}).  
      \item[]3: For each predictor $1\leq j\leq p$, compute the truncated versions $\hat{S}^{\tau_j,\tau_y}_1$, $\hat{S}^{\tau_j,\tau_y}_2$ and $\hat{S}^{\tau_j,\tau_y}_3$ and thus $\widehat{rdcov}_j$ and $\widehat{rdc}_j$ from equation (\ref{eq:1}). 
     \end{itemize}  
         \vspace{-0.2cm}
  \item[]\textbf{Screening}: Compute $\hat{\omega}^{\tau}_j=\widehat{rdc}_j^2$ and sort all features by $\hat{\omega}^{\tau}_j$ from the largest to the smallest. 
         \vspace{-0.2cm}
    \item[] \textbf{Output}: The top $d$ features ranked by $\hat{\omega}^{\tau}_j$. 
  \end{itemize} 
  \addtocounter{algorithm}{-1}
 \end{algorithm}
 
Following sure screening, one can perform a second-stage variable selection by commonly used regularization methods such as Lasso, Elastic net or SCAD, to select the final set of predictors. Alternatively, to identify the important predictors that are marginally independent of the response, we may refer to the original SIS screening method in \cite{fan2008sure} and propose an iterative version of RDC-SIS by removing variables in stepwise fashion. Due to limited space, we will not further expand its discussion here.

\section{Theoretical properties}
\label{sec:theory}

\setcounter{equation}{0} 
\section{Simulation}
\label{sec:sim}

\subsection{Simulation settings}
\label{subsec:sim_setting}

In this section, we will conduct comprehensive Monte Carlo simulations to show the performance of our proposed screening method as compared to other existing methods under different scenarios. In addition to our RDC method, we include five other popular sure screening methods for comparison, including marginal correlation screening (SIS; \cite{fan2008sure}), the original distance correlation screening (DC; \cite{li2012feature}), sure independent ranking and screening (SIRS; \cite{zhu2011model}), fused Kolmogorov filter (FK; \cite{mai2015fused}) and nonparametric independence screening (NIS; \cite{fan2011nonparametric}). For FK method, we consider $G_i = 2,\ldots,6$. For each of the following simulation scenarios, we replicate for $B=100$ times and evaluate four criteria of the screening performance below: 

\begin{itemize}
 \item Minimum model size (MMS): the minimum model size to include all active predictors. 
 \item $\mathcal{P}_1$: the proportion of replications that include at least one true active predictor for a given model size $d$. 
 \item $\mathcal{P}_a$: the proportion of replications that include all true active predictors for a given model size $d$. 
 \item $TP_d$: average number of true active predictors identified for a given model size $d$.
\end{itemize}
For $\mathcal{P}_1$ and $\mathcal{P}_a$, we set $d=\lfloor n/\log n \rfloor$. When calculating $TP_d$, we further introduce two larger model sizes $d_1=\lfloor  2n/\log n \rfloor$, $d_2=\lfloor  3n/\log n \rfloor$ for a comprehensive evaluation. 

\noindent\textbf{(I).} In the first scenario, we consider general simulation with various types of heavy-tailed distribution in $\mathbf{x} =(X_1,X_2,\ldots,X_p)^T$: 
 \begin{itemize}
\item [] Ia. $\mathbf{x} \sim t_{3}(0,\Sigma_{p\times p})$, where $\Sigma_{jj}=1$ and $\Sigma_{jj'}=0.5^{|j-j'|}$ for $1\le j\neq j'\le p$;
\item [] Ib. $X_j$ independently drawn from $Pareto(1,1)$ for $1\le j\le p$.
\end{itemize}

In all models, we consider $n=100$ and $p=2000$. We generate the response $Y \sim t_2(\mathbf{x}^T \beta)$, where the indices of nonzero $\beta$'s represent the true active predictors. Here, we set the nonzero part of $\beta$ equal to $(2,-2,2,-2)$ (i.e. the number of true active predictors $s=4$) and let them evenly distributed among all $p$ variables. We evaluate and compare the performance of different methods.


\ 

\noindent\textbf{(II).} In the second scenario, we consider the multivariate response. We generate $\mathbf{x}$ from the same heavy-tailed distributions as in scenario (I), and then generate $\mathbf{y}=(Y_1,Y_2,Y_3)^T$ from a multivariate normal distribution with common mean $\mathbf{x}^T \beta$ and $\Sigma_Y=(\sigma_{Y,kk'})_{3\times 3}$, where $\Sigma_{Y,kk}=1$ and $\Sigma_{Y,kk'}=0.5$ for $k\neq k'$. The nonzero parts of $\beta$ are $\beta_1=\beta_2=1$ (i.e. the number of true active predictors $s=2$). 


\

\noindent\textbf{(III).} In the third scenario, we consider the simulation example 1(a) and 1(b) in \cite{li2012feature}. However, instead of setting $\mathbf{x}$ to be multivariate normal distributed, we alternatively consider $\mathbf{x}$ to have multivariate central t distribution with degrees of freedom equal to 3. Specifically, we assume $\mathbf{x} =(X_1,X_2,\ldots,X_p)^T  \sim t_{3}(0,\Sigma_{p\times p})$, where $\Sigma_{jj}=1$ and $\Sigma_{jj'}=0.5^{|j-j'|}$ for $1\le j\neq j'\le p$.  We fix $n=200$ and $p=2000$ as in their examples. The responses $Y$ are generated from the following two models:  
\begin{itemize}
\item [] IIIa. $Y = \alpha_1\beta_1X_1 + \alpha_2\beta_2 X_2 + \alpha_3 \beta_3 I(X_{12}<0) + \alpha_4 \beta_4 X_{22} + \epsilon$; 
\item [] IIIb. $Y = \alpha_1\beta_1X_1X_2 + \alpha_3 \beta_2 I(X_{12}<0) + \alpha_4 \beta_3 X_{22} + \epsilon$,
\end{itemize}
where $\beta_j=(-1)^U(a+|Z|)$ for $j=1,2,3$ and $4$, where $a=4\log n/\sqrt{n}$ with $U\sim Bern(0.4)$ and $Z\sim N(0,1)$. We set $(\alpha_1,\alpha_2,\alpha_3,\alpha_4)=(2,0.5,3,2)$ as in \cite{li2012feature}. 


\subsection{Simulation results} 
\label{subsec:sim_results}

Table \ref{tab:1} shows the results of the first scenario. It is clear that RDC outperforms DC under both t distribution (Ia) and Pareto distribution (Ib). In addition, RDC is among the best performer when comparing to all other methods for both cases. Such advantage is also seen in the second scenario when the response is multivariate (Table \ref{tab:2}). 


For the third scenario (Table \ref{tab:3}), when the true regression does not include interaction terms (IIIa), the perform of all methods are very closed. In the presence of $X_1X_2$ interaction, DC as well as RDC outperform the SIS and SIRS methods, which is consistent with what has been found in \cite{li2012feature}. Because here $\mathbf{x}$ follows a heavy-tailed multivariate t distribution, it is also expected to see that RDC outperforms the original DC based screening methods.


\begin{table}
\caption{Scenario (I) results: $Y \sim t_3(\mathbf{x}^T \boldsymbol{\beta})$, $n=100$, $p=2000$, $s=4$, $\beta_s \in \{-2,2\}$, $\rho_X=0.5$, $B=100$.}
\label{tab:1}       
\begin{tabular}{cccc ccc}
\hline
Models & Method & Minimum model size & $\mathcal{P}_1$ & $\mathcal{P}_a$ & $TP_{d_1}$ & $TP_{d_2}$ \\
\hline   
  \multirow{7}{*}{Ia: $\mathbf{x} \sim t_{3}(0,\Sigma_{p\times p})$} & RDC  & 15.5  & 1    & 0.6  & 3.7  & 3.83 \\
 &  DC   & 18    & 1    & 0.53 & 3.6  & 3.77 \\
 &  SIS  & 157.5 & 0.93 & 0.1  & 2.83 & 3.07 \\
 &  SIRS & 38.5  & 1    & 0.47 & 3.53 & 3.67 \\
 &  FK   & 32.5  & 1    & 0.37 & 3.5  & 3.6  \\
 &  NIS  & 202   & 0.9  & 0.03 & 2.5  & 2.67 \\
 \hline
 \multirow{7}{*}{Ib: $ X_j \sim Pareto(1,1) $} & RDC  & 5    & 1 & 0.87 & 3.87 & 3.9  \\
 &  DC   & 43.5 & 1 & 0.2  & 3.13 & 3.53 \\
 &  Cor  & 117  & 1 & 0    & 1.97 & 2.63 \\
 &  SIRS & 16   & 1 & 0.53 & 3.67 & 3.83 \\
 &  FK   & 14   & 1 & 0.7  & 3.8  & 3.83 \\
 &  NIS  & 340  & 1 & 0.03 & 2    & 2.47 \\ 
  \hline  
\end{tabular}
\end{table}

\begin{table}
\caption{Scenario (II) results: $\mathbf{y}=(Y_1,Y_2,Y_3)^T \sim \mathbf{x}^T \beta, \Sigma_Y=(\sigma_{Y,kk'})_{3\times 3}$, $n=100$, $p=2000$, $s=2$, $\rho_X=\rho_Y=0.5$, $B=100$.}
\label{tab:2}       
\begin{tabular}{cccc ccc}
\hline
Models & Method & Minimum model size & $\mathcal{P}_1$ & $\mathcal{P}_a$ & $TP_{d_1}$ & $TP_{d_2}$ \\
\hline
  \multirow{7}{*}{IIa: $\mathbf{x} \sim t_{3}(0,\Sigma_{p\times p})$} & RDC  & 4      & 1    & 1    & 2    & 2    \\
 &  DC   & 5      & 0.97 & 0.87 & 1.93 & 1.93 \\
 &  SIS  & 1137   & 0.07 & 0    & 0.1  & 0.2  \\
 &  SIRS & 1683.5 & 0.07 & 0    & 0.07 & 0.1  \\
 &  FK   & 28.5   & 0.9  & 0.43 & 1.63 & 1.63 \\
 &  NIS  & 2      & 1    & 0.87 & 1.9  & 1.97 \\
 \hline
 \multirow{7}{*}{IIb: $ X_j \sim Pareto(1,1) $} & RDC  & 2    & 1 & 1    & 2    & 2    \\
 &  DC   & 12.5 & 1 & 0.7  & 1.87 & 1.93 \\
 &  SIS  & 41.5 & 1 & 0.43 & 1.53 & 1.67 \\
 &  SIRS & 3    & 1 & 0.83 & 1.83 & 1.87 \\
 &  FK   & 2    & 1 & 1    & 2    & 2    \\
 &  NIS  & 38   & 1 & 0.43 & 1.57 & 1.6 \\ 
  \hline  
\end{tabular}
\end{table}

\begin{table}
\caption{Scenario (III) results: $n=200$, $p=2000$, $\rho_X=0.5$, $B=100$, $d_1=\lfloor  2n/\log n \rfloor$, $d_2=\lfloor  3n/\log n \rfloor$.}
\label{tab:3}       
\begin{tabular}{cccc ccc}
\hline
Models & Method & Minimum model size & $\mathcal{P}_1$ & $\mathcal{P}_a$ & $TP_{d_1}$ & $TP_{d_2}$ \\
\hline
  \multirow{7}{*}{IIIa:\ $\mathbf{x} \sim t_3(0,\Sigma_{p\times p}) $} & RDC  & 112   & 1 & 0.23 & 3.23 & 3.3  \\
 &  DC   & 205.5 & 1 & 0.2  & 3.17 & 3.23 \\
 &  SIS  & 516   & 1 & 0.1  & 2.77 & 2.87 \\
 &  SIRS & 125.5 & 1 & 0.33 & 3.37 & 3.4  \\
 &  FK   & 132.5 & 1 & 0.33 & 3.27 & 3.33 \\
 &  NIS  & 768   & 1 & 0.1  & 2.53 & 2.53 \\ 
\hline   
\multirow{7}{*}{IIIb:\ $\mathbf{x} \sim t_3(0,\Sigma_{p\times p}) $}& RDC  & 20.5   & 1    & 0.5  & 3.63 & 3.77 \\
 &  DC   & 87.5   & 0.97 & 0.3  & 3.1  & 3.23 \\
 &  SIS  & 1440.5 & 0.43 & 0    & 0.7  & 0.77 \\
 &  SIRS & 1312   & 1    & 0    & 2    & 2    \\
 &  FK   & 132    & 0.97 & 0.17 & 2.8  & 3.03 \\
 &  NIS  & 1639.5 & 0.7  & 0    & 1.53 & 1.63 \\       
\hline
\end{tabular}
\end{table}

\section{Real data application}
\label{sec:real}

We then apply our method to a real genomic dataset from the Pancreatic adenocarcinoma (PAAD) cohort of The Cancer Genome Atlas (TCGA), a large consortium project that studies multi-platform molecular profiles spanning 33 cancer types \citep{weinstein2013cancer}. PAAD is the most common type of pancreatic cancer and has one of the poorest prognosis among all cancers. Previous studies showed that KRAS mutation and RAS-MAPK signaling pathway were essential to the understanding of pancreatic cancer \citep{ryan2014pancreatic,furukawa2015impacts,raphael2017integrated}. Our purpose is to investigate the effects of upstream gene expression on the expression of MAPK1 protein, a key component in the MAPK pathway. We downloaded the RNA sequencing (RNA-seq) gene expression data and retrieved the MAPK1 protein expression from the Reverse Phase Protein Array (RPPA) data of the TCGA PAAD cohort from Broad GDAC Firehose ({\color{blue} https://gdac.broadinstitute.org/}). RNA-seq data are measured in Transcripts Per Million (TPM) values and RPPA data contain continuous intensity values. After filtering genes with average TPM values smaller than 10 in RNA-seq and merging the available samples from both data types, 16191 genes for 116 cancer cases were left for the analysis. We will apply our robust DC method as well as its competitors to rank and identify the genes most predictive of the MAPK1 protein expression. Since we do not know the ground truth in real data, we will use pathway analysis results to validate the top ranked features kept by each screening method and evaluate their feature screening performance. For a fair comparison, we first take the top 500 genes ranked by each method and perform a pathway enrichment analysis by Fisher's exact test on each set using GO, KEGG, Reactome and Biocarta databases downloaded from MSigDB (cite). Pathways in the intersection of the top 200 enriched pathways from each method were regarded as function domains most indicative of the underlying biological truth and treated as a ``gold standard'' for comparison. We will plot the mean and standard errors of -log10(p-value) from the enrichment tests of these pathways using varying number of top ranked features in each screening method to compare the overall performance of the method. The best method will have the most enrichment in these pathways regardless of the number of features used. The predictive performance of the final set of genes selected after applying the second stage regularization is assessed by considering a robust version of mean absolute error (MAE) via the leave-one-out cross-validation: $MAE(\hat{\beta)} = \frac{1}{n_{test}}\sum\limits_{i=1}^{n_{test}} |Y_i^{test} - (\mathbf{x}^{test}_i)^T\hat{\beta}|$, where $Y_i^{test}$ and $\mathbf{x}^{test}_i$ are the response and predictor variables of the observation in the test data. In addition, to show the robustness and stability of our method, we also generate multiple subsamples of the original dataset and assess the concordance of features selected with the full data. 

\begin{figure}[!h]
\centering
\includegraphics[scale=0.45]{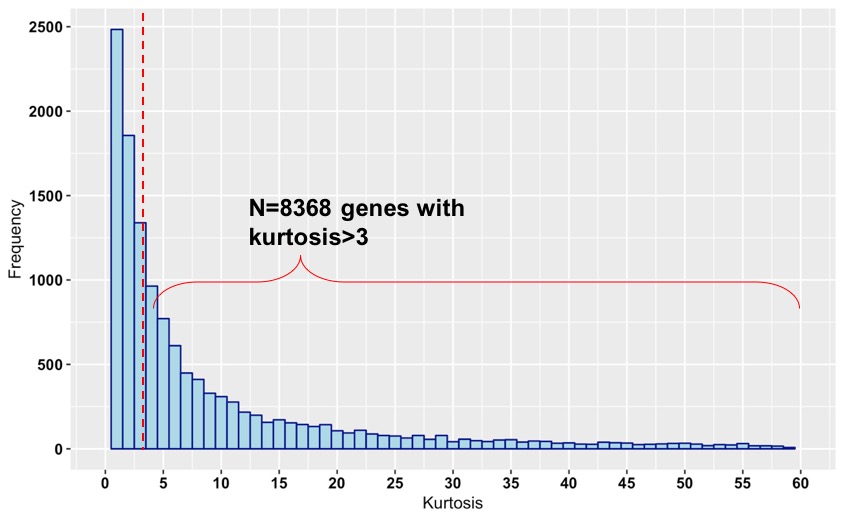}
\caption{Distribution of kurtosis of expression of all genes. The kurtosis of MAPK1 protein expression is equal to 5.5.}
\label{fig:1}
\end{figure}

\begin{figure}[!h]
\centering
\includegraphics[scale=0.45]{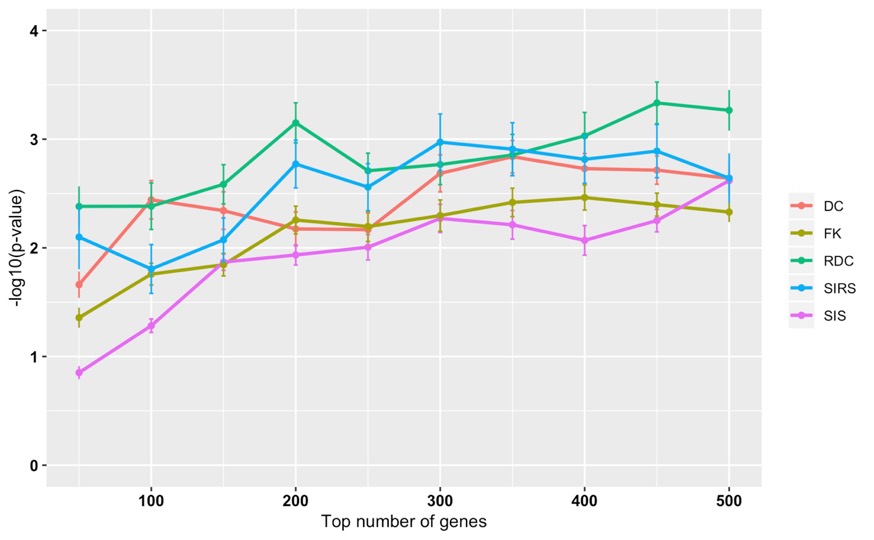}
\caption{Plot of mean -log10(p-value) with standard error from the Fisher's exact tests of the 34 pathways in the intersection against the top number of genes used for each method.}
\label{fig:2}
\end{figure}

An initial exploration of the data finds that the expression of more than half of the genes as well as the protein expression of MAPK1 show heavy tailed distribution with kurtosis larger than 3 (Figure \ref{fig:1}). A total of 34 pathways sit in the intersection of the top 200 pathways enriched by each method and will be used as the basis for comparison. These pathways mainly include interferon and other cytokine signaling, Ras and Rho signaling as well as cell cycle related pathways. Figure \ref{fig:2} shows that RDC outperforms all other methods in overall enrichment of the ``gold standard'' pathways and such trend is consistent independent of the top number of genes used. This has proven the ranking by our robust DC method is superior than the other methods by prioritizing the more informative and biologically meaningful genes. 

We keep only the top $d=\lfloor n/\log(n) \rfloor=24$ genes from each screening method and perform a second stage regularization by Lasso. The tuning parameter in Lasso is chosen by cross-validation using glmnet package \citep{friedman2009glmnet}. Table \ref{tab:5} summarizes the MAE, 
adjusted $R^2$ and selected genes of each method. Among all methods, the final model selected by RDC+Lasso has the largest adjusted $R^2$ value and has the smallest MAE. Overall, there is little overlap among the genes selected by all methods. DC and RDC shared a large proportion of matched genes (75\%) and the three genes RDC uniquely selected include $ZFP36L2$, $COQ2$ and $IFIT1$. $ZFP36L2$ has been shown to promote cancer cell aggressiveness in pancreatic ductal adenocarcinoma \citep{yonemori2017zfp}. $COQ2$ provides instructions for making an enzyme that produces coenzyme Q10, whose level in plasma has been shown to be associated with prostate cancer risk \citep{chai2011plasma}. $MAPK1$ gene itself is not selected in the top lists of any methods probably due to differences in translational rates and protein half‐lives, among others \citep{edfors2016gene}.


\begin{table}
\begin{center}
\small
\caption{Comparison of $MAE$ (computed from leave-one-out cross-validation), adjusted $R^2$ and gene lists of the final models selected by different methods}
\label{tab:5}       
\begin{tabular}{p{2cm} p{1cm} p{2cm} p{1cm} p{9cm}}
\hline
Methods & MAE & Adjusted $R^2$ & Model size & Selected genes  \\
\hline  
DC+Lasso &0.2 & 0.436 & 12 & \textit{CCNE1,CENPB,CNIH3,HAPLN1,HOXD9,IKBKB,MX1,} \\ 
 & &  & & \textit{NUDT19, PDCD2L,PDCD5,SNTA1,USP18}  \\
RDC+Lasso & 0.17 & 0.451 & 12 & \textit{CCNE1,CENPB,COQ2,HAPLN1,IFIT1,IKBKB,MX1,} \\
 & &  & & \textit{NUDT19,PDCD2L,SNTA1,USP18,ZFP36L2,ZNF204P} \\
SIS+Lasso & 0.17 & 0.394 & 8 & \textit{ATP6AP2,CCNE1,CENPB,CNIH3,FAM54A,HOXD9,} \\
 & &  & &  \textit{IMPDH1,PDCD2L} \\
FK+Lasso & 0.18 & 0.419 & 10 & \textit{ASB14,C16orf80,CCNE1,COX4NB,E2F1,HAPLN1,IKBKB,}\\ 
 & &  & & \textit{PDCD2L,SNTA1,USP18} \\
SIRS+Lasso & 0.18 & 0.419 & 12 & \textit{C16orf80,CCNE1,CENPB,CNIH3,GPI,MX1,NUDT19,NUDT2,}\\ 
 & &  & & \textit{OAS2,PDCD2L,SNTA1,ZFP36L2} \\
  \hline        
\end{tabular}
\end{center}
\end{table}

\section{Discussion}
\label{sec:discuss}

In this paper, we proposed a robust version of distance correlation measure called ``Robust DC'' to perform variable screening in ultrahigh-dimensional regression and accounted for both model misspecification and tail robustness. The method builds upon the original formulas for estimating distance correlation and replaces with its truncated version that protects against heavy-tailedness. We also proposed a data-driven procedure to determine the optimal level of truncation by automatically tuning the robustification parameter. Simulation under scenarios of various heavy-tailed distributions demonstrated the advantage of our robust DC method in prioritizing true active predictors when either the predictor or the outcome variables have heavy tails. The real application to the heavy-tailed RNA-seq and RPPA data of TCGA PAAD cohort revealed that the robust DC method identified more biologically informative pathways and selected unique genes related to the progression mechanism of prostate cancer. 

Since it was first introducted by \cite{fan2008sure}, sure screening methods based on different model-based or model-free measures have been proposed and shown their merits in variable selection of ultrahigh-dimensional regression and classification problems when coupled with regularization methods. Distance correlation based sure screening is one such representative method that often sits as one of the top performers in different comparison scenarios \citep{li2012feature}. However, like many other model-free screening methods, the method(s) are robust against model misspecification but fail to address the tail robustness. Heavy-tailedness is another important characteristics of modern big data and its great impact on mean estimation has been reiterated in recent literature \citep{fan2017estimation}. Our method is the first sure screening method to address both model misspecification and tail robustness. Importance of addressing heavy-tailedness is seen in real data where a number of critical biomarkers are not selected by other methods due to the skewed distribution of RNA-seq data. 


The data-driven tail truncation procedure for protecting against heavy-tailedness is not restricted to DC measure but can be extended to other measures such as the Pearson correlation, the canonical correlation \citep{di2021sure}, etc. In addition, since the marginal screening methods completely ignore the inter-feature dependency, one can also consider conducting a robust multivariate (or joint) screening that ranks each variable after controlling for other correlated variables \citep{he2019covariance}. We applied the proposed screening method to the gene expression data in real data demonstration, considering the heavy-tailedness of most RNA-seq or microarray-type data. The proposed method can be readily extended to select variables in other heavy-tailed data types such neuroimaging or copy number variation data, etc.


\vskip 14pt
\noindent {\large\bf Supplementary Materials}

The supplementary materials contain the proofs of the three lemmas. 

\par
\vskip 14pt
\noindent {\large\bf Acknowledgments}


\markboth{\hfill{\footnotesize\rm TIANZHOU MA AND ZHAO REN} \hfill}
{\hfill {\footnotesize\rm Tianzhou Ma and Zhao Ren} \hfill}

\bibhang=1.7pc
\bibsep=2pt
\fontsize{9}{14pt plus.8pt minus .6pt}\selectfont
\renewcommand\bibname{\large \bf References}
\expandafter\ifx\csname
natexlab\endcsname\relax\def\natexlab#1{#1}\fi
\expandafter\ifx\csname url\endcsname\relax
  \def\url#1{\texttt{#1}}\fi
\expandafter\ifx\csname urlprefix\endcsname\relax\def\urlprefix{URL}\fi


\bibliographystyle{apalike}
\bibliography{Rdc}


\vskip .65cm
\noindent
Tianzhou Ma \\
\noindent
Department of Epidemiology and Biostatistics, University of Maryland, College Park, MD 20742. \\
\noindent
E-mail: tma0929@umd.edu
\vskip 5pt

\noindent
Zhao Ren \\
\noindent
Department of Statistics, University of Pittsburgh, Pittsburgh, PA 15261. \\
\noindent
E-mail: zren@pitt.edu
\vskip 5pt

\end{document}